\documentclass[aps,prl,twocolumn,groupedaddress,longbibliography]{revtex4-2}
\usepackage{graphicx}
\usepackage[version=3]{mhchem}
\usepackage{booktabs}
\usepackage{hyperref}

\begin{document}

\title{Reactivity of the Si(100)-2$\times$1-Cl surface with respect to PH$_3$, PCl$_3$, and BCl$_3$: Comparison with PH$_3$ on Si(100)-2$\times$1-H}

 \author{T. V. Pavlova$^{1, 2}$}
 \email{pavlova@kapella.gpi.ru}
 \author{K. N. Eltsov$^{1}$}
 \affiliation{$^{1}$Prokhorov General Physics Institute of the Russian Academy of Sciences, Moscow, Russia}
 \affiliation{$^{2}$HSE University, Moscow, Russia}

\begin{abstract}

Despite the interest in a chlorine monolayer on Si(100) as an alternative to hydrogen resist for atomic-precision doping, little is known about its interaction with dopant-containing molecules. We used the density functional theory to evaluate whether a chlorine monolayer on Si(100) is suitable as a resist for \ce{PH3}, \ce{PCl3}, and \ce{BCl3} molecules. We calculated reaction pathways for  \ce{PH3}, \ce{PCl3}, and \ce{BCl3} adsorption on a bare and Cl-terminated Si(100)-2$\times$1 surface, as well as for \ce{PH3} adsorption on H-terminated Si(100)-2$\times$1, which is widely used in current technologies for atomically precise doping of Si(100) with phosphorus. It was found that the Si(100)-2$\times$1-Cl surface has a higher reactivity towards phosphine than Si(100)-2$\times$1-H, and, therefore, unpatterned areas are less protected from undesirable incorporation of \ce{PH3} fragments. On the contrary, the resistance of the Si(100)-2$\times$1-Cl surface against the chlorine-containing molecules turned out to be very high. Several factors influencing reactivity are discussed. The results reveal that phosphorus and boron trichlorides are well-suited for doping a patterned Cl-resist by donors and acceptors, respectively.

\end{abstract}

\maketitle

\section{Introduction}

The interaction of dopant-containing molecules with the Si(100) surface is important for functionalization reactions used in many practical applications \cite{2010Perrine}. In particular, the fabrication of nanoelectronic devices on single atoms requires functionalization of the Si(100) surface with atomic precision \cite{2012Fuechsle, 2019He}. To create silicon-dopant bonds in preselected areas, the resist patterned by scanning tunneling microscope (STM) lithography is dosing with dopant-containing molecules \cite{2001OBrien, 2003Schofield}. One of the most important characteristics of an atomic resist is its low reactivity with respect to dopant-containing molecules \cite{2007Goh}.

Currently, a hydrogen monolayer is used as a resist on the Si(100) surface, and phosphine serves as a donor-containing molecule \cite{2012Fuechsle, 2019He, 2001OBrien, 2003Schofield, 2007Goh}. Recently, a chlorine monolayer has been proposed as an alternative resist \cite{2018Pavlova}. Although STM-lithography on the Si(100)-2$\times$1-Cl surface has been demonstrated \cite{2019Dwyer, 2020Pavlova}, the selectivity of the phosphine reaction with a chlorinated surface has been only estimated \cite{2018Pavlova}. Besides, it can be reasonable to use trichlorides as dopant-containing molecules with Cl-resist (for example, \ce{BCl3} \cite{2020Silva-Quinones, 2021Dwyer}, \ce{PCl3}) instead of trihydrides. However, it has previously been reported that the chlorinated surface is highly reactive towards \ce{BCl3} \cite{2020Silva-Quinones} and \ce{NH3} \cite{2011Soria}, with activation barriers for ammonia incorporation even lower than that on the bare surface. At the same time, a chlorine monolayer was successfully used as a resist for \ce{BCl3} \cite{2021Dwyer}, and the calculated activation barriers for the reaction of \ce{NH3} with a chlorinated surface were quite high \cite{2008Lange}. Thus, the reactivity of the Cl-terminated Si(100) surface with respect to some dopant-containing molecules is still unclear.

This work is aimed to establish the possibility of using chlorine as a resist for \ce{PH3}, \ce{PCl3}, and \ce{BCl3} by studying the selectivity of the reactions of such molecules with passivated and depassivated Si(100)-2$\times$1-Cl. We used density functional theory (DFT) to calculate reaction pathways for \ce{PH3}, \ce{PCl3}, and \ce{BCl3} incorporation into the Cl-terminated Si(100)-2$\times$1 surface. As a reference, the reaction of \ce{PH3} with the H-terminated Si(100)-2$\times$1 surface was considered. To ensure that the selectivity of the studied reactions is high, we also calculated the adsorption and dissociation of the \ce{PH3}, \ce{PCl3}, and \ce{BCl3} molecules on the bare Si(100) surface and in H and Cl vacancies. The reaction pathways were considered with the removal of excess atoms into the gas phase as well as with their insertion into the surface. We found that \ce{PCl3} and \ce{BCl3} exhibit very high selectivity towards the reactions with the Si(100)-2$\times$1-Cl surface with depassivated areas.

\section{Calculation details}

The spin-polarized DFT calculations were performed using the projector augmented wave method implemented in Vienna \textit{Ab-initio} Simulation Package (VASP) \cite{1993Kresse, 1996Kresse}. For the exchange-correlation potential, generalized gradient approximation (GGA) by Perdew, Burke, and Ernzerhof (PBE) \cite{1996Perdew} was employed. To account van der Waals corrections we used DFT-D2 method developed by Grimme \cite{2006Grimme}. The kinetic-energy cutoff of the plane wave basis was set up to 400\,eV. The silicon surfaces were modeled by an eight layer slabs consisting of 4$\times$4 supercell and a vacuum region of about 13\,{\AA}. Chlorine or hydrogen atoms were placed on the upper face of the slabs to form a Si(100)-2$\times$1 structure, whereas H atoms saturate the dangling bonds at the bottom face of the slabs. The bottom two Si layers were frozen at their bulk position, while the coordinates of other atoms were fully relaxed until the residual forces were smaller than 0.01\,eV/\,{\AA}. Brillouin zone integrations were done using a 4$\times$4$\times$1 k-point grid.

The calculated Si lattice constant (5.415\,{\AA}) is in a good agreement with the experimental value (5.416\,{\AA}). We used GGA-PBE with DFT-D2 because it is a good compromise between accuracy and computational cost to study the surface reactions considered in this work. For the accurate energy evaluation of the reaction of \ce{PH3}, \ce{PCl3}, and \ce{BCl3} molecules with H- and Cl-terminated Si(100)-2$\times$1, the size of the supercell is important. A larger surface supercell allows the better relaxation of the chlorine layer upon the trichloride molecule incorporation, and sixty layers in a slab are required to accurately describe the electronic structure of the Si(100)-2$\times$1 surface \cite{2017Sagisaka}. However, we have limited the size of the supercell to 4$\times$4$\times$8 to enable calculations of reaction barriers in a reasonable time.

The activation barriers were calculated by using the nudged elastic band (NEB) method \cite{1998NEB} and further validated by the climbing image nudged elastic band (CI-NEB) method \cite{2000CNEB}. In this case, the criterion of convergence for residual forces was reduced to 0.03\,eV/\,{\AA}. The number of images (including the two end points) in the NEB method ranged from seven for simple adsorption pathways to twelve for complex pathways. For further refinement calculation by the CI-NEB method, three to seven images were used, one of which was climbed to the highest saddle point.

\section{Results}

This section presents the results of \ce{PH3} reaction with H- and Cl-terminated Si(100)-2$\times$1 surfaces, as well as \ce{PCl3} and \ce{BCl3} reaction with Cl-terminated Si(100)-2$\times$1. On the Si(100)-2$\times$1 surface covered with a monolayer of hydrogen (chlorine), each dangling bond of a Si dimer terminates with H (Cl) \cite{1995Waltenburg}. \ce{PH3} and \ce{PCl3} are trigonal pyramidal molecules with a lone pair of electrons, while \ce{BCl3} is a planar trigonal molecule with an empty lone pair. In the initial states, the molecules were located 5\,{\AA} above the surface, with the P atom closer to the surface than the H (Cl) atoms of the \ce{PH3} (\ce{PCl3}) molecule. The adsorption energy of the initial state was considered as the sum of the energies of the surface and the molecule in the gas phase. In the most favorable physisorbed states, \ce{PH3} and \ce{PCl3} molecules are located above hydrogen (chlorine) monolayer in the position between the H (Cl) atoms of the adjacent four dimers and have approximately the same adsorption energies, $-0.25$ eV for \ce{PH3} on Si(100)-2$\times$1-H, $-0.16$ eV for \ce{PH3} on Si(100)-2$\times$1-Cl, and $-0.24$\,eV for \ce{PCl3} on Si(100)-2$\times$1-Cl. The energy of the physisorbed state of \ce{BCl3} on Si(100)-2$\times$1-Cl is nearly the same ($-0.21$ eV), but the molecule is located above the dimer row.

At the beginning of the section, we present the results on the adsorption of \ce{PH3}, \ce{PCl3}, and \ce{BCl3} on a clean Si(100) surface and the molecular adsorption in a single vacancy on the passivated surfaces: \ce{PH3} in H and Cl vacancies, and \ce{PCl3} and \ce{BCl3} in a Cl vacancy.

\subsection{Adsorption of PH$_3$, PCl$_3$, and BCl$_3$ on clean Si(100)}

\begin{figure}[h]
 \includegraphics[width=\linewidth]{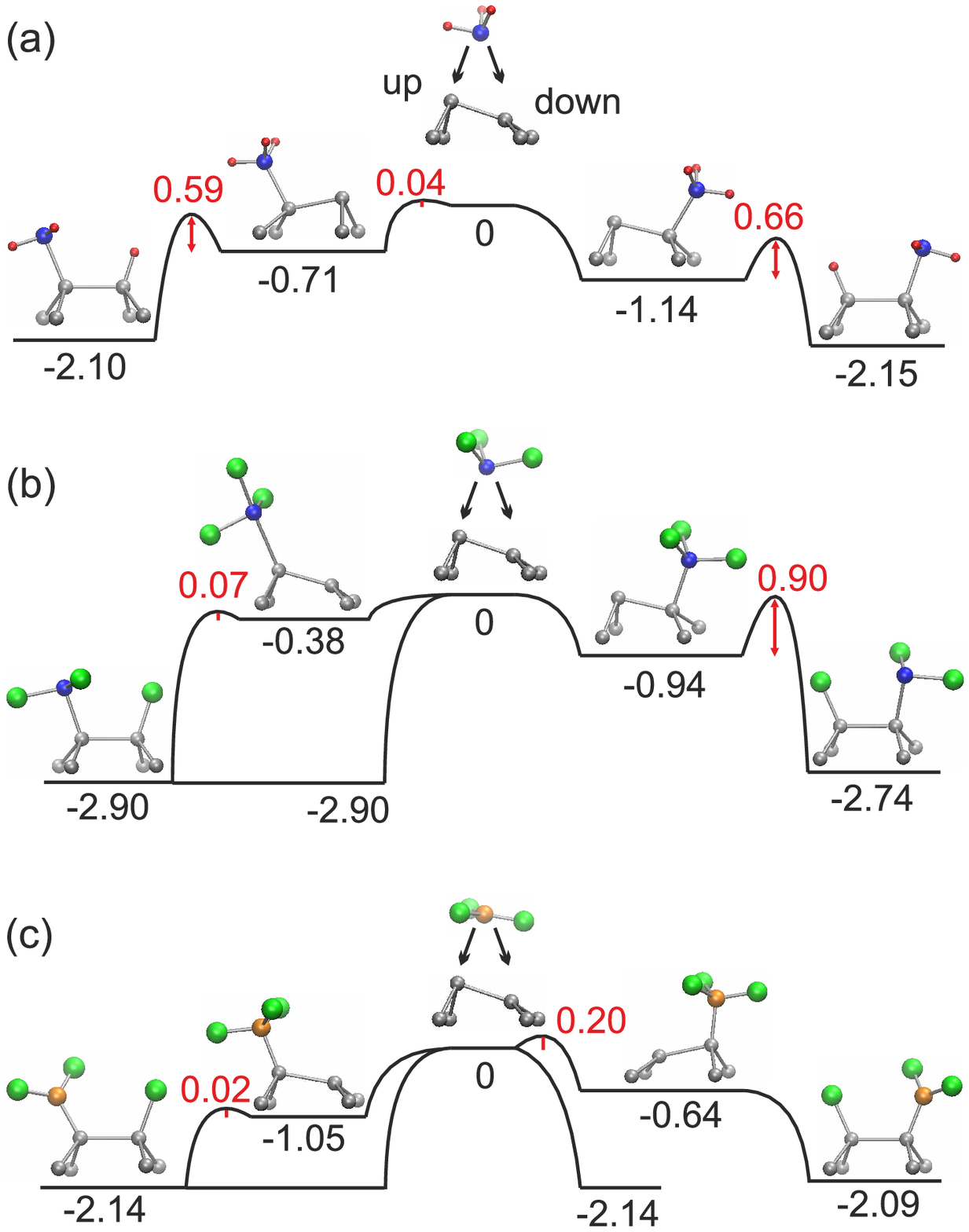}
\caption{\label{clean} Energy barrier diagrams of adsorption and the first stage of dissociation of \ce{PH3} (a), \ce{PCl3} (b), and \ce{BCl3} (c) on the clean Si(100) surface. Black numbers indicate adsorption energies, red numbers indicate activation barriers. All energies are given in eV. Silicon atoms are marked in gray, hydrogen in red, chlorine in green, phosphorus in blue, and boron in orange.}
\end{figure}

This section presents the results for the \ce{PH3}, \ce{PCl3}, and \ce{BCl3} adsorption on a clean Si(100) surface with the p(2$\times$2) reconstruction \cite{1995Waltenburg}. On the Si(100) surface, the lower Si atom of the dimer known as the `down atom' is electrophilic, and it donates an electronic charge to the nucleophilic  `up atom' \cite{1997Konecny}. Figure~\ref{clean} shows pathways of \ce{PH3}, \ce{PCl3}, and \ce{BCl3} adsorption on the clean Si(100) surface and the first stage of dissociation with the loss of hydrogen (chlorine) on the same Si dimer. The preferred adsorption site of all three molecules is the silicon dangling bond, as was previously found for phosphine \cite{1994Cao, 2005Warschkow}. Two different paths of the \ce{PH3}, \ce{PCl3}, and \ce{BCl3} adsorption on the down and up atom are shown in Fig.~\ref{clean} to the right and left of the initial state, respectively.

Phosphine adsorption on the Si(100) surface has been extensively studied  \cite{2001Miotto, 2005Warschkow, 2005McDonell, 2016Warschkow}, and it was found that \ce{PH3} bound to the electrophilic (down) silicon atom by the lone pair of electrons (Fig.~\ref{clean}a). From this local minimum, phosphine dissociates into \ce{PH2} and H fragments. Our calculations of this reaction path are consistent with the results of previous works, although the activation barrier of dissociation (0.66 eV) is larger than the barrier calculated in the cluster model (0.46 eV) \cite{2016Warschkow}. After \ce{PH3} adsorption on the up atom, we found that the tilt angle of the dimer changed and \ce{PH3} again binds to the down atom (Fig.~\ref{clean}a). This adsorption position is less favorable due to the change in the dimer slope relative to the adjacent dimers, in agreement with the previous study \cite{2005Warschkow}.

The \ce{PCl3} molecule also prefers to adsorb on the down atom (Fig.~\ref{clean}b). Further dissociation of \ce{PCl3} proceeds with a higher activation barrier (0.90 eV) than in the case of \ce{PH3}. In addition, we found two reaction paths at the \ce{PCl3} adsorption on the up atom, depending on the molecule orientation. If one of the Cl atoms is above the down Si atom, a Si-Cl bond is formed, followed by spontaneous dissociation of the \ce{PCl3} molecule. Otherwise, \ce{PCl3} adsorbs on the up atom without an activation barrier and dissociates into \ce{PCl2} and Cl with a very low activation barrier (0.07 eV).

The \ce{BCl3} molecule is preferentially adsorbed on the nucleophilic up atom (Fig.~\ref{clean}c) since \ce{BCl3} has an empty lone pair of electrons. If \ce{BCl3} approaches the surface so that the B atom is above the down atom, the slope of the dimer is reversed, which requires activation energy of 0.20 eV. As a result, the \ce{BCl3} molecule forms a Si-B bond with the up atom. If \ce{BCl3} approaches the surface so that one Cl atom interacts with the down atom, dissociation into \ce{BCl2} and Cl occur spontaneously. In contrast to the \ce{PH2}-Si and \ce{PCl2}-Si surface units, all three bonds of the B atom of the \ce{BCl2}-Si unit lie in the same plane \cite{2009Ferguson}.

Summarizing the results of this section, the \ce{PH3} and \ce{PCl3} molecules prefer to adsorb on an electrophilic atom, whereas \ce{BCl3} prefers to adsorb on a nucleophilic atom. The most stable configurations for all chemisorbed molecules have close energies, about $-1$ eV. After dissociation, the energy of the surface structure with the \ce{PCl2} unit becomes lower than that of \ce{PH2} and \ce{BCl2}, by about 0.8 eV. The pathways for adsorption and dissociation depend on the molecule orientations. In contrast to phosphine, phosphorus and boron trichlorides can dissociate spontaneously on the clean Si(100) surface.

\subsection{Molecular adsorption in H and Cl vacancies}

The structure with a phosphine adsorbed in a H vacancy on the Si(100)-2$\times$1-H surface is significantly less stable (adsorption energy $-0.32$ eV) than that on the clean surface ($-1.14$ eV). A stronger P-Si bond is formed on the clean surface due to the electrophilic nature of the down Si atom. Interestingly, the structure with \ce{PH3} adsorbed in a Cl vacancy ($-0.87$ eV) is much more stable than in a H vacancy. This can be explained by the electronegativity of chlorine, which can lead to a more electrophilic Si atom on the Cl-terminated Si(100)-2$\times$1 surface than on the H-terminated one.

The adsorption of \ce{PCl3} in a Cl vacancy on Si(100)-2$\times$1-Cl ($-0.37$ eV) is less preferable than on the clean surface ($-0.94$ eV). The adsorption of \ce{BCl3} in a Cl vacancy with the formation of a Si-Cl ($-0.25$ eV) or Si-B ($+0.09$ eV) bond is also much less favorable than on the clean surface ($-1.05$ eV). We attribute this to a strong repulsive interaction between the Cl atoms of the surface and the Cl atoms of the molecule.

Thus, in all cases, the molecules adsorbed in a single vacancy on the H- or Cl-terminated Si(100)-2$\times$1 surface are weaker bound to a Si atom than in the case of the clean surface. The adsorption in vacancies proceeded without any barriers, in agreement with the previous calculation for the \ce{PH3} adsorption in a Cl vacancy \cite{2018Pavlova}.

\subsection{Reaction of PH$_3$ with H-terminated Si(100)-2$\times$1}

\begin{figure*}[t]
\includegraphics[width=\linewidth]{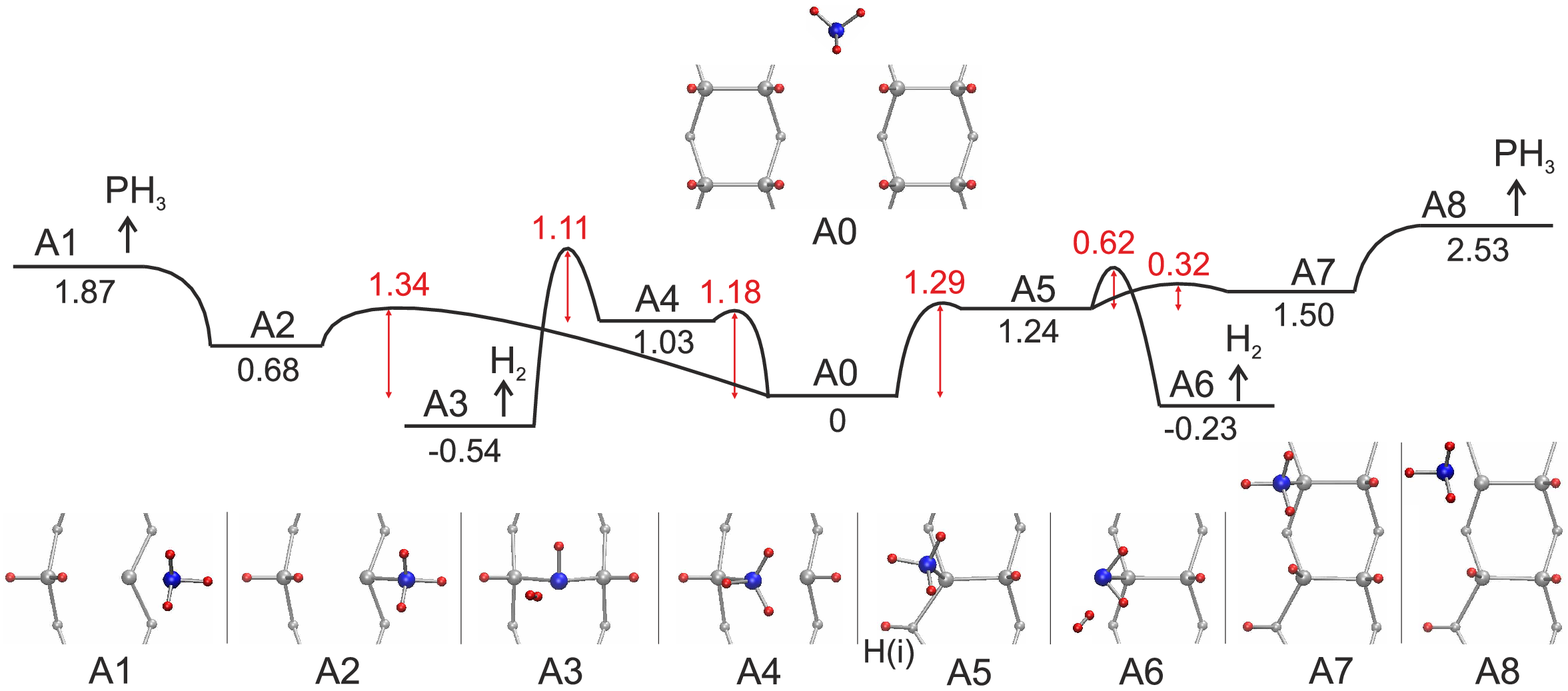}
\caption{\label{SiH_PH3} Energy barrier diagrams of \ce{PH3} adsorption on the Si(100)-2$\times$1-H surface. Pathways start at the center from the initial state A0. A top view is shown for each surface structure A0--A8. Silicon atoms are marked in gray, hydrogen in red, and phosphorus in blue. Si atoms of the upper layer are indicated by large gray circles, while Si atoms of the second layer are indicated by small ones. All energies are given in eV relative to the initial state energy. Red numbers indicate activation barriers.}
\end{figure*}

Energy barrier diagram of \ce{PH3} insertion into the H-terminated Si(100)-2$\times$1 surface is shown in Fig.~\ref{SiH_PH3}. On the right-hand side of the reaction pathway relative to the initial configuration A0, the \ce{PH3} molecule substitutes a H atom by pushing it into the groove between the dimer rows (configuration A5 in Fig.~\ref{SiH_PH3}). The H(i) atom inserted in the groove between the dimer rows is attached to the second-layer Si atom \cite{2020PavlovaPCCP}. Hydrogen desorption from configuration A5 lowers the energy of the system, resulting in the configuration with the \ce{PH2} unit on the surface (A6). If the H(i) atom diffuses in the groove (A7), it cannot combine with a H atom of \ce{PH3} and desorb. As a result, \ce{PH3} bound with a Si atom will remain on the surface. After desorption of  \ce{PH3} from the A7 configuration, two defects are left on the surface: the Si dangling bond and the H(i) atom. The energy of the resulting surface structure with the defects (A8) is 2.53\,eV higher than that of the initial perfect surface (A0).

On the left-hand side of the reaction pathway, all H atoms remain bonded to first-layer Si atoms upon the phosphine incorporation. In configuration A4, the \ce{PH3} molecule is adsorbed on the Si atom occupied by the H atom, which leads to the breaking of the bond inside the Si dimer. After overcoming an energy barrier of about 1.1 eV for \ce{H2} desorption, we arrive at the lowest-energy configuration with the PH unit bonded to two Si atoms (A3). In another configuration, \ce{PH3} is adsorbed on the Si atom, and a dihydride is formed on the same dimer (A2). Phosphine desorption from configuration A2 leads to a significant increase in energy since a dihydride and a dangling bond remain on the surface (A1).

In summary, as the result of the \ce{PH3} interaction with the H-terminated Si(100)-2$\times$1 surface, the \ce{PH3}, \ce{PH2}, and PH units can be formed. As can be seen in Fig.~\ref{SiH_PH3}, the activation barriers of \ce{PH3} adsorption are much higher than the barriers of the reverse reactions; therefore, the probability of finding \ce{PH3} on the surface in configurations A2, A4, and A5 is low. However, \ce{PH3} can be bound to the surface in configuration A7, since the migration of the H(i) atom makes the reverse reaction unlikely. The incorporation of the \ce{PH2} (A6) and PH (A3) units is accompanied by the release of \ce{H2}, and the configuration A3 is more favorable than the initial configuration A0. The activation barrier is lower at the \ce{PH2} formation (A6), but the surface structure with PH is more stable (A3). Phosphine desorption at the end of the reaction path leaves defects on the surface, which is highly unfavorable (A1, A8).

\subsection{Reaction of PH$_3$ with Cl-terminated Si(100)-2$\times$1}

\begin{figure*}[t]
 \includegraphics[width=\linewidth]{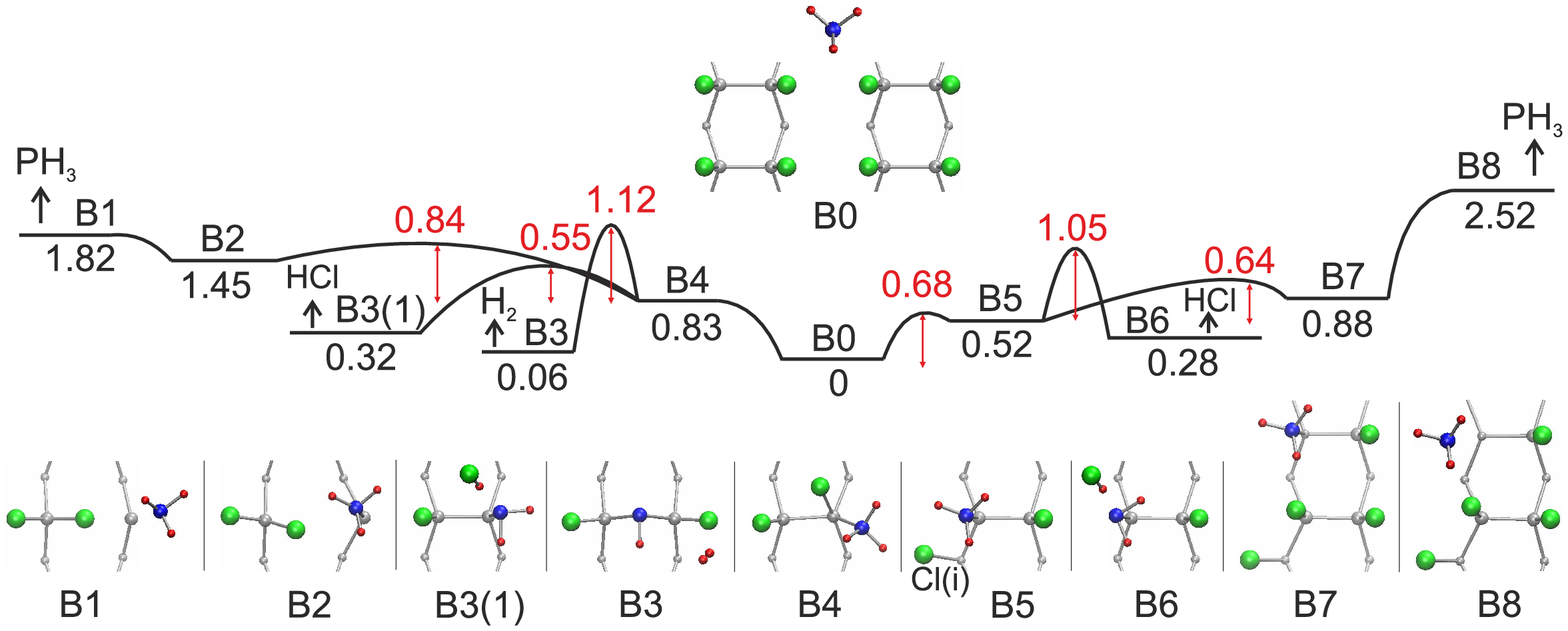}
\caption{\label{SiCl_PH3} Energy barrier diagrams of \ce{PH3} adsorption on the Si(100)-2$\times$1-Cl surface. Pathways start at the center  from the initial state B0. A top view is shown for each surface structure B0--B8. Silicon atoms are marked in gray, hydrogen in red, chlorine in green, and phosphorus in blue. Si atoms of the upper layer are indicated by large gray circles, while Si atoms of the second layer are indicated by small ones. All energies are given in eV relative to the initial state energy. Red numbers indicate activation barriers.}
\end{figure*}

Figure~\ref{SiCl_PH3} shows the pathways of phosphine chemisorption on the Cl-terminated Si(100)-2$\times$1 surface, which are accompanied by the displacement of one Cl atom inside the Si dimer (B4) or into the groove between dimer rows (B5). A Cl(i) atom inserted in the groove forms a bond with a second-layer Si atom \cite{2020PavlovaPRB}, like a H(i) atom. The Cl(i) atom can combine with the H atom of the \ce{PH3} unit and desorb as HCl (B6). If the Cl(i) atom migrates in the groove, the \ce{PH3} unit will remain on the surface (B7). Finally, \ce{PH3} can desorb (B8), but this is the least stable state along the reaction path since surface defects such as Cl(i) and a dangling bond are formed. This path is very similar to that for the H-terminated Si(100)-2$\times$1 surface, but the most stable state along this path with the \ce{PH2} unit (B6) is formed with a lower barrier (1.57 eV). Note that here and below, the reaction barrier was calculated as the maximum energy that must be overcome from the initial state, i.e., the sum of the energy of the intermediate configuration B5 (0.52 eV) and the activation barrier B5$\rightarrow$B6 (1.05 eV).

The left side of the \ce{PH3} reaction pathway on a chlorinated surface is also similar to that of a hydrogenated one, except that \ce{PH3} can not fit inside a chlorinated dimer, as in the case of a hydrogenated dimer (A4). Instead, the \ce{PH3} molecule shifts the Cl atom and binds to the Si atom (B4) with the subsequent formation of the \ce{PH2} unit with HCl desorption (B3(1)) or the PH unit with \ce{H2} desorption (B3). Configuration B3 is the most stable among all the considered configurations, but configuration B3(1) requires minimal activation energy (1.38 eV). The reaction with \ce{PH3} desorption (B1) after a dichloride formation (B2) is similar to that with a dihydride formation (A2$\rightarrow$A1), both in energy and in activation barrier.

Thus, upon the interaction of phosphine with Si(100)-2$\times$1-Cl, the PH, \ce{PH2}, and \ce{PH3} units can be formed. In the most stable configuration (0.06 eV), the incorporation of the PH unit is accompanied by \ce{H2} desorption (B3). The incorporation of the PH  and \ce{PH2} units into a chlorinated surface occurs with lower activation barriers than into a hydrogenated one, but the final states are less stable. At the same time, the formation of surface defects on a chlorinated surface requires approximately the same energy as in the case of a hydrogenated one. Thereby, according to our calculations, the reactivity of the chlorinated surface with respect to \ce{PH3} is higher than that of the hydrogenated one. To reduce the undesirable incorporation of the phosphorus-containing molecule into the Cl-terminated Si(100)-2$\times$1 surface, we suggest to use the \ce{PCl3} molecule instead of \ce{PH3}.

\subsection{Reaction of PCl$_3$ with Cl-terminated Si(100)-2$\times$1}

\begin{figure*}[t]
 \includegraphics[width=0.83\linewidth]{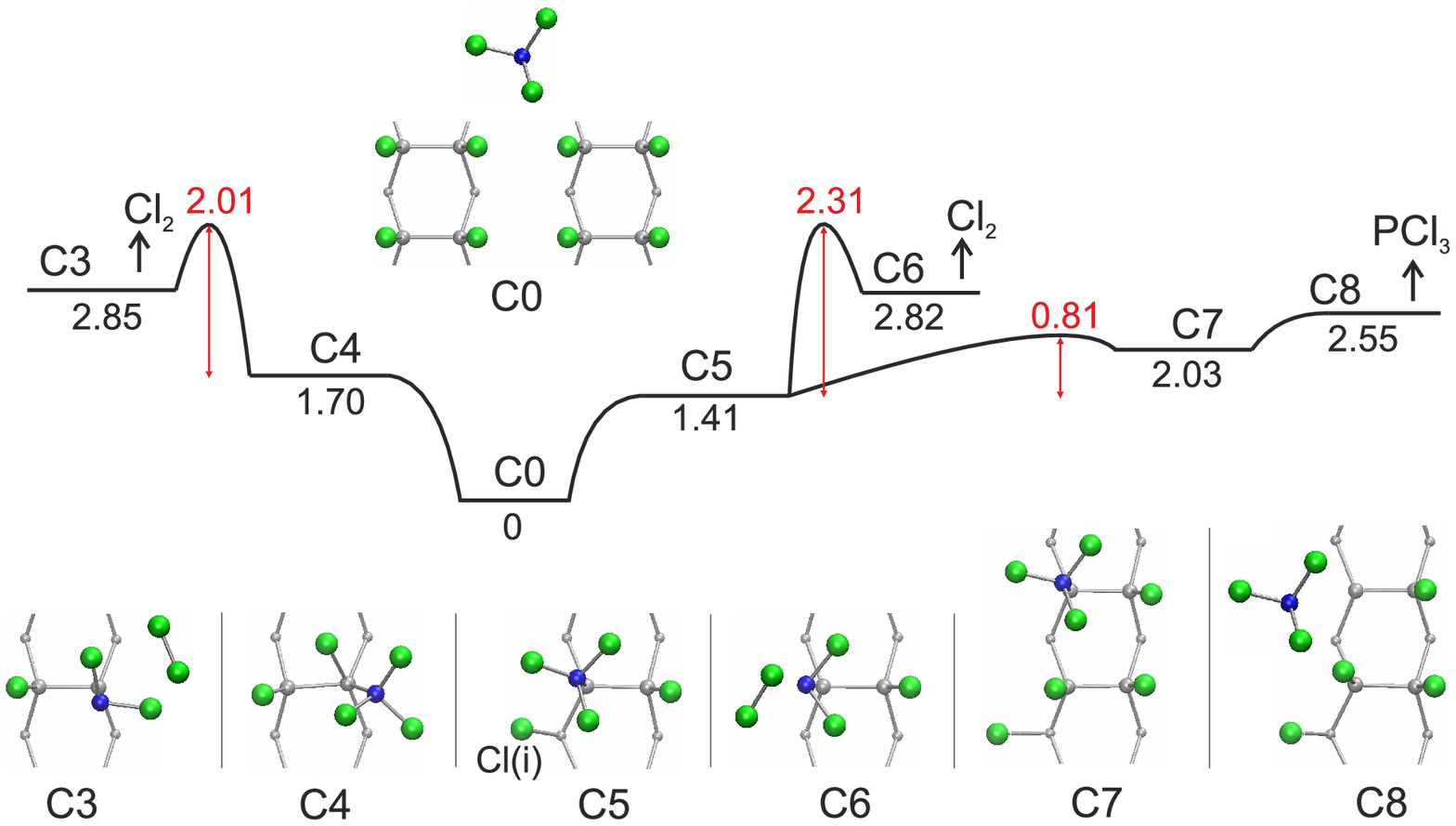}
\caption{\label{SiCl_PCl3} Energy barrier diagrams of \ce{PCl3} adsorption on the Si(100)-2$\times$1-Cl surface. Pathways start at the center from the initial state C0. A top view is shown for each surface structure C0, C3--C8. Silicon atoms are marked in gray, chlorine in green, phosphorus in blue. Si atoms of the upper layer are indicated by large gray circles, while Si atoms of the second layer are indicated by small ones. All energies are given in eV relative to the initial state energy. Red numbers indicate activation barriers.}
\end{figure*}

The reaction pathway of the \ce{PCl3} molecule on the Cl-terminated Si(100)-2$\times$1 surface is shown in Fig.~\ref{SiCl_PCl3}. Again, the surface structures along the path with Cl(i) inserted in the groove (to the left from the initial state C0) are similar to those for \ce{PH3} on chlorinated and hydrogenated surfaces. However, the most stable configuration along this path for \ce{PH3} (A6) is disfavored for \ce{PCl3} (C6). The incorporation of the \ce{PCl2} unit accompanied by \ce{Cl2} desorption (C6) has very high energy (2.82 eV) and activation barrier (3.72 eV).

Another possible way to incorporate \ce{PCl3} is to shift a Cl atom and form a bond with a Si atom (C4). Note that in the case of \ce{PCl3} adsorption, we did not find any stable configurations like A2, B2, or A3 in the case of \ce{PH3} adsorption. After \ce{Cl2} desorption (C3), the \ce{PCl2} unit can be formed on the surface. Thus, upon \ce{PCl3} adsorption on the Cl-terminated Si(100)-2$\times$1 surface, the final configurations with the \ce{PCl2} or \ce{PCl3} units have significantly higher adsorption and activation energies than similar configurations with \ce{PH2} or \ce{PH3}. On the other hand, the energy of the final states with surface defects formed after \ce{PCl3} desorption is approximately the same as in the above reactions of phosphine with the H- and Cl-terminated Si(100)-2$\times$1 surfaces.

\subsection{Reaction of BCl$_3$ with Cl-terminated Si(100)-2$\times$1}

\begin{figure*}[t]
 \includegraphics[width=\linewidth]{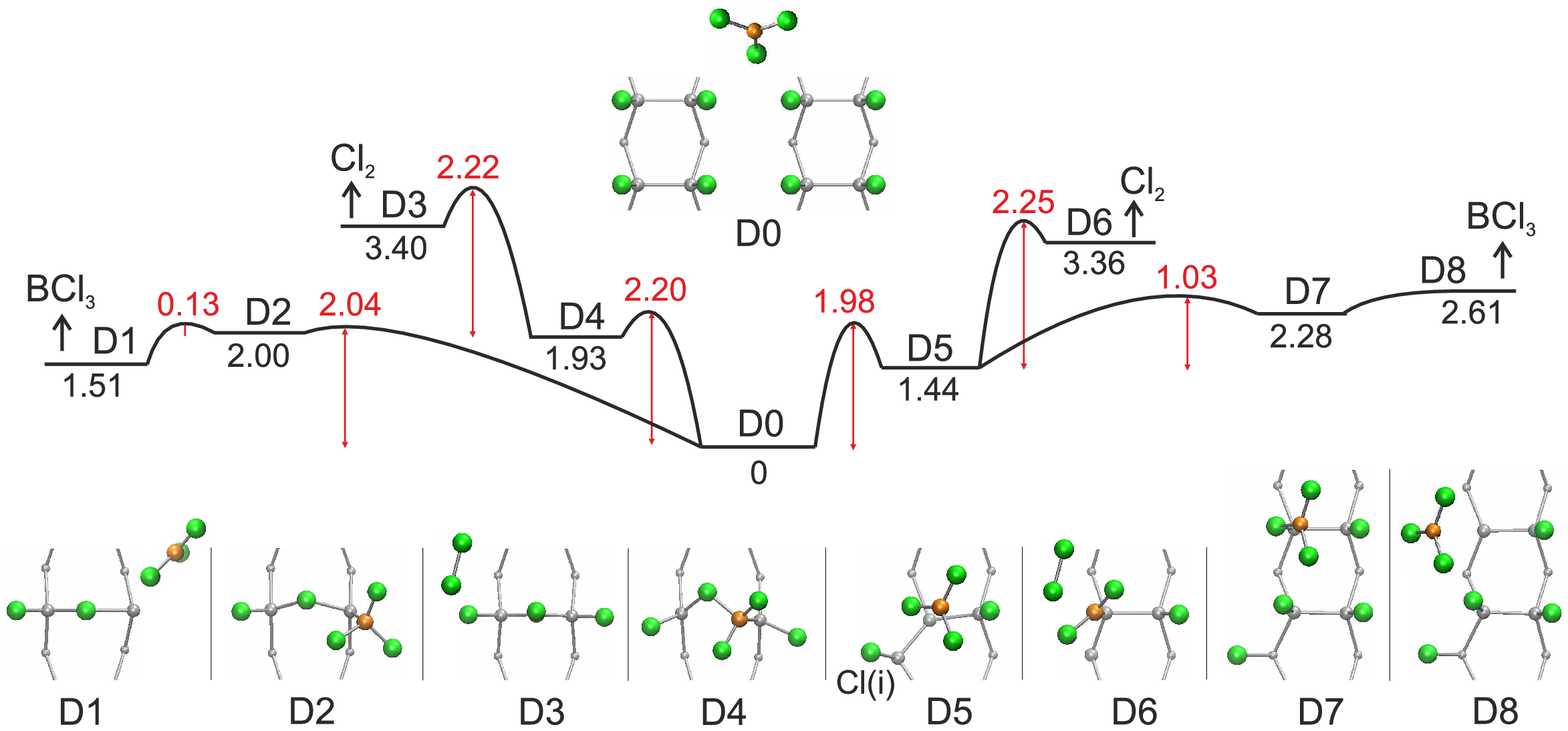}
\caption{\label{SiCl_BCl3} Energy barrier diagrams of \ce{BCl3} adsorption on the Si(100)-2$\times$1-Cl surface. Pathways start at the center from the initial state D0. A top view is shown for each surface structure D0--D8. Silicon atoms are marked in gray, chlorine in green, boron in orange. Si atoms of the upper layer are indicated by large gray circles, while Si atoms of the second layer are indicated by small ones. All energies are given in eV relative to the initial state energy. Red numbers indicate activation barriers.}
\end{figure*}

Figure~\ref{SiCl_BCl3} illustrates the pathways of \ce{BCl3} adsorption on the Cl-terminated Si(100)-2$\times$1 surface. In the D5 configuration, the distance from the chlorine atom of the \ce{BCl3} molecule to the nearest Si atom exceeds 4\,{\AA}.  After overcoming an energy barrier of 3.69 eV, the chlorine molecule can desorb, and the Si-B bond is formed on the surface (D6). If Cl(i) migrates in the groove, the \ce{BCl3} molecule approaches the surface, forming a Si-Cl bond (D7). However, this Si-Cl bond is rather weak since the \ce{BCl3} molecule can easily desorb (D8).

The \ce{BCl3} molecule can be adsorbed on the Si atom shifting the Cl atom to the bridge position between two Si atoms (D2), and after that \ce{BCl3} can desorb leaving Cl in the bridge position (D1). This path (D0$\rightarrow$D1) requires minimum activation energy (2.13 eV), but as a result, no B-Si bond is formed on the surface. Alternatively, \ce{BCl3} can be adsorbed on the Si atom shifting the Cl atom to the bridge position between the B and Si atoms (D4). From configuration D4, we calculated \ce{Cl2} desorption with the SiCl-BCl-SiCl fragment formation on the surface, in which the B atom is located in the middle of the dimer under the central Cl atom (D3). According to the calculated paths, the B-Si bonds remain on the surface only in the D3 and D6 configurations. Note that we did not find any stable configuration with the \ce{BCl3} molecule (as well as with \ce{PH3} and \ce{PCl3}) attached to the second-layer Si atom.

\section{Discussion}

To enable the utilizing of the Cl-terminated Si(100)-2$\times$1 surface as a resist, direct silicon-dopant bonds must be formed in the depassivated areas only. On the bare Si(100) surface, dissociative adsorption of \ce{PCl3} and \ce{BCl3} can occur spontaneously, but if \ce{PH3}, \ce{PCl3}, and \ce{BCl3} are molecularly adsorbed, dissociation into two fragments requires activation energy (Fig.~\ref{clean}). The reaction barrier for \ce{PH3} dissociation into \ce{PH2} and H is easily overcome at room temperature \cite{2016Warschkow}. Consequently, the adsorbed \ce{BCl3} molecule will rapidly dissociate at room temperature since it has a lower activation barrier. Taking into account that the time scale of \ce{PH3} dissociation is of the order of 10 $\mathrm{\mu s}$  \cite{2016Warschkow}, the adsorbed \ce{PCl3} molecule is also expected to be short-lived at room temperature because the activation barrier for \ce{PCl3} dissociation is only slightly higher than for \ce{PH3}. Thus, on the bare Si(100) surface, the first stage of  \ce{PH3}, \ce{PCl3}, and \ce{BCl3} dissociation into two fragments will proceed at room temperature.

To compare the resistance of a hydrogen and chlorine monolayer to the direct Si-dopant bond formation, we put together the lowest activation energies for all the considered reactions in Table~\ref{table1}. At phosphine adsorption on the chlorinated surface, the \ce{PH2} unit formation requires an activation energy of 1.38 eV, which is approximately 0.5 eV lower than that for the \ce{PH2} unit on the hydrogenated surface. Thus, the Si(100)-2$\times$1-Cl surface is more reactive to phosphine than Si(100)-2$\times$1-H. However, the replacement of \ce{PH3} with \ce{PCl3} (\ce{BCl3}) makes the Cl-terminated Si(100)-2$\times$1 surface more protective against Si-P (Si-B) bond formation. Indeed, the activation barrier for the \ce{PCl2} and \ce{BCl2} units insertion into Si(100)-2$\times$1-Cl increases to 3.7 eV. Thus, if \ce{PCl3} and \ce{BCl3} are used as dopant-containing molecules, the chlorine monolayer protects the silicon surface from Si-P and Si-B bond formation even better than the hydrogen monolayer from Si-P bond formation at \ce{PH3} adsorption. Note that the chlorine monolayer also protects the Si(100)-2$\times$1 surface against undesirable dopant incorporation in the presence of single vacancies since molecular adsorption in vacancies is less favorable than on a clean surface.

\begin{table*}[t]
\begin{center}
\caption{Minimum energy barriers (E$_b$) for the incorporation of PH$_2$, PCl$_2$, and BCl$_2$ into the Si(100)-2$\times$1-H and Si(100)-2$\times$1-Cl surfaces together with adsorption energies (E$_a$) in these states.}
    \label{table1}
\begin{tabular}{lccrccl}
\toprule
Surface  & Molecule & E$_b$, eV & E$_a$, eV & Inserted unit  & Desorbed specie & Configuration \\
\midrule
    Si(100)-2$\times$1-H  & \ce{PH3}  & 1.86 & $-$0.23 & \ce{PH2} & \ce{H2} & A6 (Fig.~\ref{SiH_PH3}) \\
    Si(100)-2$\times$1-Cl & \ce{PH3}  & 1.38 & 0.32 & \ce{PH2} & \ce{HCl} & B3(1) (Fig.~\ref{SiCl_PH3})     \\
    Si(100)-2$\times$1-Cl & \ce{PCl3} & 3.71 & 2.85 & \ce{PCl2} & \ce{Cl2} & C3 (Fig.~\ref{SiCl_PCl3})     \\
    Si(100)-2$\times$1-Cl & \ce{BCl3} & 3.69 & 3.36 & \ce{BCl2}& \ce{Cl2} & D6 (Fig.~\ref{SiCl_BCl3})   \\
\bottomrule
\end{tabular}

\end{center}
\end{table*}

There are several factors that govern the reactivity of the Si(100)-2$\times$1-H and Si(100)-2$\times$1-Cl surfaces with respect to the \ce{PH3}, \ce{PCl3}, and \ce{BCl3} molecules. One such factor is the energy loss at the \ce{H2}, HCl, and \ce{Cl2} species desorption in the course of the molecular fragments insertion. To compare the desorption energies of these species, a 4$\times$4 supercell was prepared with one row of H-terminated dimers and another row of Cl-terminated dimers. We sequentially removed chlorine and hydrogen from the surface to form \ce{H2}, HCl, or \ce{Cl2}  in the vacuum gap between the slabs, keeping the distance between vacancies on the surface as large as possible. The desorption energies of \ce{H2}, HCl, and \ce{Cl2} obtained in this way are 2.71 eV, 3.04 eV and 5.57 eV, respectively. A comparison of the calculated desorption energies predicts that the reaction accompanied by \ce{Cl2} removal from the surface is significantly less favorable than the reaction with \ce{H2} or HCl desorption. Note that excess atoms can remain on the surface rather than desorb, but the energy also increases due to defects formation. We can therefore conclude that the undesirable incorporation of the dopant-containing molecules with chlorine (such as  \ce{PCl3} and \ce{BCl3}) into the Cl-terminated surface is very unlikely due to the high energy cost for \ce{Cl2} desorption or surface defect formation.

The steric effect is another factor that determines the lower reactivity of chlorine monolayer with respect to Cl-containing molecules compared to the reactivity of the hydrogen monolayer with respect to phosphine. Indeed, the calculated distance between the nearest Cl atoms on the Si(100)-2$\times$1-Cl surface (3.83\,{\AA}) is close to the double van der Waals radius (ionic radius) of chlorine, which ranges from 3.5 to 3.8\,{\AA} \cite{2001Batsanov}. In addition, the \ce{PCl3} and \ce{BCl3} molecules are larger than \ce{PH3}, as can be seen from the calculated P-Cl, B-Cl, and P-H bond lengths equal to 2.07\,{\AA}, 1.75\,{\AA}, and 1.43\,{\AA}, respectively. For these reasons, when we placed the \ce{PCl3} and \ce{BCl3} molecules on Si(100)-2$\times$1-Cl, in many cases they were pushed out of the surface by the repulsive forces between chlorine. Thus, the larger radius of the resist atoms, together with the size of the molecule, reduces the reactivity.

One more factor affecting reactivity is the charge distribution on the surface and in the molecules. Chlorine on the Si(100)-2$\times$1 surface takes more charge from the Si atom than hydrogen \cite{2010Perrine} due to stronger electronegativity of Cl, which implies that Si atoms under a chlorine monolayer are more electrophilic than under a hydrogen monolayer. The adsorption of phosphorus-containing molecules is facilitated by a more electrophilic Si atom, since the \ce{PH3} and \ce{PCl3} molecules have a lone pair of electrons, which makes the chlorinated surface more reactive towards these molecules. On the contrary, the \ce{BCl3} molecule is not polar and prefers to form a bond with a nucleophilic silicon atom. Consequently, the adsorption structure with \ce{BCl3} should have a lower stability than that with \ce{PCl3}, in accordance with our calculations (Table~\ref{table1}).

Recently, it was experimentally demonstrated that \ce{BCl3} reacts with the Cl-terminated Si(100) surface more effectively than with H-terminated Si(100) producing the direct Si-B bonds at temperatures even below 70$^{\circ}$C \cite{2020Silva-Quinones}. Our calculations predict a low reactivity of the Si(100)-2$\times$1-Cl surface with respect to \ce{BCl3} and therefore seem to contradict this experiment. The contradiction can be explained by dichloride species and possible defects presented on the surfaces prepared by the wet chemistry method (as was mentioned in Ref. \cite{2020Silva-Quinones}), which should have different reactivities than monochloride. In another recent work, the Si(100) surface covered with chlorine monolayer, which was verified by imaging the Cl-terminated Si(100) surface with an STM, was used to selectively deposit \ce{BCl3} \cite{2021Dwyer}. Experimental results obtained at room temperature demonstrate that the reaction proceeds selectively in the depassivated areas, in accordance with our calculations.

The selectivity of the reaction of phosphine with the lithographic patterned H-resist has been experimentally proved in the STM study \cite{2007Goh}. After adsorption at room temperature, \ce{PH3} fragments were found only within the depassivated areas. Therefore, \ce{PCl3} and \ce{BCl3} are also suitable for doping patterned Si(100)-2$\times$1-Cl surface since the activation barriers for their insertion under Cl monolayer are higher than in the case of \ce{PH3} on Si(100)-2$\times$1-H (Table~\ref{table1}). Moreover, high activation barriers ensure that the chlorine monolayer will protect Si(100) from \ce{PCl3} and \ce{BCl3} even at higher temperatures, close to the temperature of chlorine desorption from Si(100)-2$\times$1-Cl. In contrast, the Cl-terminated Si(100)-2$\times$1 surface is more reactive to phosphine than H-terminated Si(100)-2$\times$1, but this does not necessarily mean that the Cl-resist is not suitable for use with phosphine. Indeed, the activation barrier of phosphine reaction with Cl-terminated silicon (1.38 eV) is higher than with the bare surface (0.66 eV), thus the chlorine monolayer can, in principle, protect Si(100) from undesirable incorporation of \ce{PH3} at certain parameters (temperature and phosphine dose).

\section{Summary and conclusions}

In summary, DFT calculations have been performed to study the selectivity of \ce{PH3}, \ce{PCl3}, and \ce{BCl3} reactions with the Cl-terminated Si(100)-2$\times$1 surface and \ce{PH3} reaction with H-terminated Si(100)-2$\times$1. Several factors have been identified that influenced the reactivity of the considered reactions. We found that the Cl-terminated Si(100)-2$\times$1 surface is more reactive to phosphine than H-terminated Si(100)-2$\times$1. The Si(100)-2$\times$1-Cl surface is the least reactive towards \ce{PCl3} and \ce{BCl3} molecules, which allows it to use as a resist even at elevated temperatures. The reasons for the low reactivity are the steric effect and the significantly less favorable configurations with desorbed \ce{Cl2} compared to that with \ce{H2} or \ce{HCl}. Preferred adsorption of \ce{BCl3} on the nucleophilic rather than electrophilic Si atom explains the lower stability of the adsorption structures with \ce{BCl3} compared to \ce{PCl3}. Thus, the \ce{PCl3} molecules are well-suited for doping a selectively patterned Si(100)-2$\times$1-Cl surface with donors, and \ce{BCl3} with acceptors.

\section{Acknowledgments}
This study was supported by the Russian Science Foundation under the grant No. 21-12-00299. We also thank the Joint Supercomputer Center of RAS for providing the computing power.

\bibliography{Pavlova_TEX_revised.bbl}

\end{document}